\documentclass[a4paper,11pt]{fullverllncs}
\usepackage[left=2.5cm,top=2.5cm,right=2.5cm,centering]{geometry}

\usepackage{graphicx}

\usepackage{amsmath,amssymb}
\usepackage{graphicx}
\usepackage{ascmac}
\usepackage{array}
\usepackage{algpseudocode}
\usepackage{amsfonts}
\usepackage{amssymb}
\usepackage{amsmath}
\usepackage{algorithm, algpseudocode}
\usepackage{multirow}
\usepackage{arydshln}
\usepackage{hhline} 
\usepackage[misc,geometry]{ifsym} 

\usepackage[dvipdfmx, colorlinks]{hyperref, xcolor}
\definecolor{winered}{rgb}{0.5,0,0}
\definecolor{darkblue}{rgb}{0,0,0.5}
\definecolor{darkgreen}{rgb}{0,0.3,0}
\hypersetup{
linkcolor=winered,
citecolor=darkblue,
urlcolor=darkgreen
}

\newcommand{\preRO}{{${\bf prefix\mathchar`-RO}^h$}}

\newcommand{\bpreRO}{{$\overline{{\bf prefix\mathchar`-RO}^h}$}}

\newcommand{\CPCROM}{{${\rm common\mathchar`-CP\mathchar`-CT\mathchar`-ROM}$}}
\newcommand{\SPCROM}{{${\rm  CP\mathchar`-CT\mathchar`-ROM}$}}
\newcommand{\SPSPROM}{{${\rm  CP\mathchar`-SPT\mathchar`-ROM}$}}
\newcommand{\SPFPROM}{{${\rm  CP\mathchar`-FPT\mathchar`-ROM}$}}
\newcommand{\CPCO}{{${\bf CP\mathchar`-CO}^h$}}
\newcommand{\bCPCO}{{$\overline{{\bf CP\mathchar`-CO}^h}$}}

\newcommand{\Gen}{{\sf {Gen}}}
\newcommand{\Sign}{{\sf {Sign}}}
\newcommand{\Verify}{{\sf {Verify}}}
\newcommand{\RSAGen}{{\sf {RSAGen}}}

\newcommand{\GrGen}{{\sf {GrGen}}}

\spnewtheorem{assumption}{Assumption}{\bfseries}{\itshape}

\begin{document}

\title{Weakened Random Oracle Models with Target Prefix\thanks{This is a full version of a paper \cite{TYT18} in Innovative Security Solutions for Information Technology and Communications - 11th International Conference (SecITC 2018).}}

\author{Masayuki Tezuka\textsuperscript{(\Letter)} \and Yusuke Yoshida \and Keisuke Tanaka}
\authorrunning{M. Tezuka et al.}
\institute{Tokyo Institute of Technology, Tokyo, Japan\\
\email{tezuka.m.ac@m.titech.ac.jp}}

\maketitle 
\pagestyle{plain}
             
\begin{abstract}
Weakened random oracle models (WROMs) are variants of the random oracle model (ROM). 
The WROMs have the random oracle and the additional oracle which breaks some property of a hash function.
Analyzing the security of cryptographic schemes in WROMs, we can specify the property of a hash function on which the security of cryptographic schemes depends.

Liskov (SAC 2006) proposed WROMs and later Numayama et al. (PKC 2008) formalized them as CT-ROM, SPT-ROM, and FPT-ROM. In each model, there is the additional oracle to break collision resistance, second preimage resistance, preimage resistance respectively.
Tan and Wong (ACISP 2012) proposed the generalized FPT-ROM (GFPT-ROM) which intended to capture the chosen prefix collision attack suggested by Stevens  et al. (EUROCRYPT 2007). 

In this paper, in order to analyze the security of cryptographic schemes more precisely, we formalize GFPT-ROM and propose additional three WROMs which capture the chosen prefix collision attack and its variants. 
In particular, we focus on signature schemes such as {\sf RSA-FDH}, its variants, and {\sf DSA}, in order to understand essential roles of WROMs in their security proofs.

\end{abstract}


\section{Introduction}
\subsection{Background}
A hash function is an important primitive for many cryptographic schemes.
The security of these cryptographic schemes is often proved in the random oracle model (ROM) \cite{BR93}.
In the ROM, a hash function is regarded as an ideal random function. 
Instead of computing a hash value, all the parties can query the random oracle ${\cal RO}$ to get the hash value.
Compared to the standard model, it is easier to design efficient cryptographic schemes with the provable security.

In an implementation of a cryptographic scheme, the random oracle is replaced by a hash function. 
However, when a hash function is attacked, it is not clear a problem whether the security of the cryptographic scheme proven in the ROM is guaranteed.

To explain this fact, we consider two signature scheme RSA-Full-Domain-Hash ({\sf RSA-FDH}) \cite{BR96} and {\sf RSA-PFDH} \cite{Cor02} which is a variant of {\sf RSA-FDH}.
Both schemes satisfies the existential unforgeability against chosen message attacks (EUF-CMA) security in the ROM.
Now, we give an observation of the EUF-CMA security of these scheme where these scheme are implemented by a hash function $h$ whose collision resistance property is broken.

In {\sf RSA-FDH}, if a collision $(m, m ')$ satisfying $h (m) = h (m') \land m \neq m '$ is found, then the adversary makes a signing query on a message $m$, gets a valid signature $\sigma = h(m)^d \mod N$ where $d$ is a signing key and $N$ is a RSA modulo, the adversary can generate a valid forgery $(m ', \sigma)$.
This example shows that the EUF-CMA security of {\sf RSA-FDH} depends on the collision resistance property of a hash function.

By contrast, it seems that the attack mentioned above does not works for {\sf RSA-PFDH}.
In {\sf RSA-PFDH}, a signature $\sigma$ on a message $m$ is computed as $\sigma = (r, x)$ where $r$ is randomly chosen by the signer and $x = h(m||r)^d \mod N$ where $m||r$ is the concatenation of two strings $m$ and $r$.
Since randomness $r$ is chosen by a signer, even if a collision $(m||r, m'||r')$ satisfying $h (m||r) = h (m'||r') \land m||r \neq m'||r'$ is found, the adversary seldom obtains a signature $\sigma$ on a message $m$ formed as $\sigma = (m||r)$ by making a signing query on a message $m$.
Intuitively, the EUF-CMA security of {\sf RSA-PFDH} still holds even if the collision resistance property of the hash function is broken.
However, we cannot prove this fact in the ROM.

\subsubsection*{Weakened Random Oracle Models.}
Liskov \cite{Lis06} introduced the idea of weakened random oracle models (WROMs). 
Each model has the additional oracle that breaks the specific property of a hash function.
For instance, assuming that the oracle which returns the collision is added, we can capture the situation where the collision is found.

Pasini and Vaudenay \cite{PV07} used the Liskov's idea to consider the security of Hash-and-Sign signature schemes in the random oracle model with the additional oracle which returns the first preimage.

Numayama, Isshiki, Tanaka \cite{NIT08} formalized the Liskov's idea as three types of WROMs: The collision tractable ROM (CT-ROM), the second preimage tractable ROM (SPT-ROM), and the first preimage tractable ROM (FPT-ROM). Each model has the additional oracle ${\cal CO}$, ${\cal SPO}$, ${\cal FPO}$, respectively:
\begin{itemize}
\item[-] ${\cal CO}()$: It picks $x$ randomly and uniformly returns a collision $(x, x')$ such that $x \neq x'$ and $h(x) = h(x')$.
\item[-] ${\cal SPO}(x)$: Given an input $x$, it uniformly returns $x'$ such that $x \neq x'$ and $h(x) = h(x')$.
\item[-] ${\cal FPO}(y)$: Given an input $y$, it uniformly returns $x$ such that $h(x)=y$.
\end{itemize}
Since Liskov's models considered only compression functions, these oracles are different from that of Liskov's model in following respects. 
${\cal CO}$ does not provide a collision if there is no $x'$ such that $x \neq x'$ and $h(x) = h(x')$ for $x$. 
${\cal SPO}$ (resp. ${\cal FPO}$) does not provide $x'$ (resp. $x$) if there is no $x'$ (resp. $x$) such that $h(x) = h(x')$ (resp. $h(x) = y$). 

Numayama et al. analyzed the EUF-CMA security of {\sf RSA-FDH} and its variant in WROMs.
There result showed that {\sf RSA-PFDH} is EUF-CMA secure in CT-ROM.
As described before, we give intuition that the EUF-CMA security of {\sf RSA-PFDH} still holds even if the collision resistance property of the hash function is broken but we cannot prove this fact in the ROM.
Thus, by using WROMs, we can prove the security of cryptographic schemes when a hash function is attacked.

Kawachi, Numayama, Tanaka, and Xagawa \cite{KNTX10} analyzed the indistinguishability against adaptive chosen ciphertext attacks (IND-CCA2) security of {\sf RSA-OAEP} \cite{BR94}, the Fujisaki-Okamoto conversion ({\sf FO})  \cite{FO99} and its variants in WROMs. 
They showed {\sf RSA-OAEP} encryption scheme is IND-CCA2 secure even in the FPT-ROM. 

\subsubsection*{Chosen Prefix Collision Attack.}
Stevens, Lenstra and de Weger \cite{SLW07} proposed the chosen prefix collision attack for a hash function.
In this attack, we decide a pair $(P, P')$ of prefixes beforehand and find a collision $(P||S, P'||S')$.
Using this attack against {\sf MD5} \cite{Riv92}, an adversary can find a target collision by making roughly $2^{49}$ calls to the internal compression function. 

Moreover, Stevens, Sotirov, Appelbaum, Lenstra, Molnar, Osvik and de Weger \cite{SSALMOW09} succeeded in reducing the number of calls to the internal compression function to find a collision to $2^{16}$ by setting $P = P'$.
However, this attack was not captured by WROMs by Numayama et al.

To capture the chosen prefix attack, Tan and Wong \cite{TW12} proposed the generalized FPT-ROM (GFPT-ROM).
This model has the additional oracle ${\cal GFPO}$.
\begin{itemize}
\item[-] ${\cal GFPO}(y,r)$: Given an input $(y, r)$, it uniformly returns $x=m||r$ such that $h(m||r)=y$.
\end{itemize}  
They showed {\sf RSA-PFDH}$^\oplus$ \cite{NIT08} is not EUF-CMA secure in the GFPT-ROM. 
Moreover, they proposed a generic transformation of Hash-and-Sign signature schemes. If the original scheme is secure in the ROM, the converted scheme is secure in the GFPT-ROM. They proposed {\sf RSA-FDH}$^+$ by using this transformation for {\sf RSA-FDH}.

\subsection{Our Contributions.}
Thanks to transformation proposed by Tan and Wong \cite{TW12}, constructing a secure scheme in WROMs is not a serious problem.
But the security against the chosen prefix collision attack for standard signature schemes have not been clarified.
Moreover, GFPT-ROM captures a strong variant of the chosen prefix collision attack than that of attack proposed by Stevens et al and existing WROMs do not exactly capture the chosen prefix collision attack by Stevens et al.
Furthermore, like the work by Tan and Wong, we can consider other variants of the collision attack. 

In this work, we introduce new WROMs which captures the chosen prefix collision attack and its variants.
Then, we analyze the EUF-CMA security of standard signature schemes against chosen prefix collision attacks by using our WROMs. 
Our analysis of these schemes in WROMs provides a more precise security indication against chosen prefix collision attacks. 

\subsubsection*{WROMs for the Chosen Prefix Collision Attack.}
In order to analyze the security against chosen prefix collision attacks in more detail, we extend the idea of \cite{TW12} and obtain new models.
Consequently, we obtain the {\CPCROM} and {\SPCROM} from the CT-ROM. 
We also obtain the {\SPSPROM} from the SPT-ROM.
The {\CPCROM} captures the case of $P=P'$ in the chosen prefix collision attack, {\SPCROM} captures the chosen prefix collision attack, and {\SPSPROM} captures a variation of this attack that we decide $(P||S, P')$ beforehand and find a collision $(P||S, P'||S')$. 

\subsubsection*{Security Analysis in WROMs.}
We analyze {\sf RSA-FDH} and its variants in our new WROMs. 
The analysis results in these WROMs are given in Fig. \ref{summarize1}. 
In this table, models become weaker as it goes right of the table. 
The security in a weaker model indicates the scheme is secure against stronger attacks to hash functions.
Our result indicates {\sf RSA-PFDH} and {\sf RSA-PFDH}$^\oplus$ are secure in the {\SPCROM}, but not secure in the {\SPSPROM}. 
Surprisingly, the analysis result of {\sf RSA-PFDH}$^\oplus$ is interesting in that even if a cryptographic scheme is secure in the SPT-ROM or FPT-ROM, it may not be secure in the {\SPSPROM}. 

In 2018, Jager, Kakvi and May \cite{JKM18} prove that {\sf RSASSA-PKCS-v1.5} \cite{PKCS} is EUF-CMA secure in the ROM under the RSA assumption.
{\sf RSASSA-PKCS-v1.5} and {\sf DSA}  \cite{Ker13} are not analyzed using WROMs in previous works \cite{NIT08,TW12}.
We show that both signature schemes are not EUF-CMA secure in both the CT-ROM and {\CPCROM}.

\subsection{Related Works}
Unruh \cite{Unr07} proposed the ROM with oracle-independent auxiliary inputs.
In this model, an adversary $\mathcal{A}$ consists of ($\mathcal{A}_{1}$,  $\mathcal{A}_{2}$).  
The first step, computationally unbounded $\mathcal{A}_{1}$ can full access to ${\cal RO}$ and store information (e.g collisions)  in the string $z$.  
In the second step, $z$ is passed to the bounded running time adversary $\mathcal{A}_{2}$ who can access to ${\cal RO}$. 
Unruh showed that {\sf RSA-OAEP} encryption scheme is IND-CCA2 secure in the ROM with oracle-independent auxiliary inputs.

One may think that the ROM with oracle-independent auxiliary inputs already emcompasses WROMs, but it is not clear.
In the ROM with oracle-independent auxiliary inputs, $\mathcal{A}_{2}$ cannot query additional oracles that breaks hash functions after given the public key. 
By contrast, in WROMs, an adversary can query additional oracles that breaks hash functions after given the public key.
Situations captured by WROMs and the ROM with oracle-independent auxiliary inputs are different and relevance between WROMs and the ROM with oracle-independent auxiliary inputs is not still clear.
In particular, even if a cryptographic scheme is insecure in the ROM with oracle-independent auxiliary inputs, it is not clear whether it is secure in WROMs or not.
Hence it is worth analyzing cryptographic schemes with WROMs.

\begin{figure*}[htbp]
  \begin{center}
     \begin{tabular}{|c||c|c;{1pt/1pt}c|c|c|} \hline
     &  & \multicolumn{2}{c|}{CT-ROM \cite{NIT08}} & \multicolumn{1}{c|}{SPT-ROM \cite{NIT08}} & \multicolumn{1}{c|}{FPT-ROM \cite{NIT08}}  \\ \cdashline{3-6}[1pt/1pt]
     
   Schemes\textbackslash Models & \multicolumn{1}{c|}ROM & \multicolumn{1}{c;{1pt/1pt}}{${\sf common\mathchar`-CP\mathchar`-}$} & \multicolumn{1}{c|}{${\sf CP\mathchar`-CT\mathchar`-}$} & \multicolumn{1}{c|}{${\sf CP\mathchar`-SPT\mathchar`-}$} & \multicolumn{1}{c|}{${\sf CP\mathchar`-FPT\mathchar`-}$} \\   
    
        &  & \multicolumn{1}{c;{1pt/1pt}}{CT-ROM Def \ref{CP-CT-ROM}} & \multicolumn{1}{c|}{ROM Def \ref{SP-CT-ROM}} & \multicolumn{1}{c|}{ROM Def \ref{SP-SPT-ROM}} & \multicolumn{1}{c|}{ROM \cite{TW12}}  \\  \hline
         
    {\sf RSA-FDH} & $\checkmark$ &\multicolumn{2}{c|} {\hspace{15pt} $\times$   \cite{NIT08}} & $\times$ & $\times$     \\ \cdashline{3-6}[1pt/1pt]
    \cite{BR96} &  \cite{BR96}  & \hspace{20pt} $\times$ Th \ref{FDH-CP-CT}& $\times$ & $\times$ & $\times$  \\ \hline 
       
    {\sf RSA-PFDH} & \multirow{2}{*}{$\checkmark$}  &\multicolumn{2}{c|}  {\hspace{15pt} $\checkmark$   \cite{NIT08}} &  \hspace{15pt} $\times$  \cite{NIT08}  & $\times$ \\   \cdashline{3-6}[1pt/1pt]
    \cite{Cor02} &  & $\checkmark $  & \hspace{20pt} $\checkmark $  Th \ref{PFDH-SP-CT}  & \hspace{20pt} $\times$ Th \ref{PFDH-SP-SPT}   & $\times$   \\ \hline 
    {\sf RSA-PFDH}$^\oplus$ & \multirow{2}{*}{$\checkmark$}  &\multicolumn{2}{c|}  {$\checkmark$}    & $\checkmark$    & \hspace{15pt} $\checkmark$  \cite{NIT08}  \\ \cdashline{3-6}[1pt/1pt]
    \cite{NIT08} &  & $\checkmark$  & \hspace{20pt} $\checkmark$ Th \ref{PFDHo+-SP-CT} & \hspace{20pt} $\times$ Th \ref{PFDHo+-SP-SPT} & \hspace{15pt} $\times$ \cite{TW12}   \\ \hline
    {\sf RSA-FDH}$^+$ & \multirow{2}{*}{$\checkmark$}  & \multicolumn{2}{c|}  {$\checkmark$}    & $\checkmark$    & \hspace{15pt} $\checkmark$  \cite{TW12}  \\ \cdashline{3-6}[1pt/1pt]
    \cite{TW12} &  & $\checkmark$  & $\checkmark$    & $\checkmark$    & \hspace{15pt} $\checkmark$  \cite{TW12}  \\ \hline
    {\sf RSASSA-}& $\checkmark$& \multicolumn{2}{c|}  {\hspace{20pt} $\times$ Th \ref{RSASSAPKCS-CT}} & $\times$ & $\times$  \\   \cdashline{3-6}[1pt/1pt]    
    {\sf PKCS-v1.5} \cite{PKCS} & \cite{JKM18}  & \hspace{20pt} $\times$ Th \ref{RSASSAPKCS-CP-CT}  & $\times$ & $\times$ & $\times$  \\ \hline
    {\sf DSA} & \multirow{2}{*}{$?^*$} & \multicolumn{2}{c|} {\hspace{20pt} $\times$ Th \ref{DSA-CT}} & $\times$ & $\times$  \\   \cdashline{3-6}[1pt/1pt]
    \cite{Ker13}& & \hspace{20pt} $\times$  Th \ref{DSA-CP-CT} & $\times$ & $\times$ & $\times$  \\ \hline
    \end{tabular}
    \label{tab:const}
  \end{center}
 \caption{The EUF-CMA security of signature schemes}\label{summarize1}
Def: Definition, Th: Theorem, $\checkmark$: secure, $\times$: insecure, $?$: security proof has not been provided. \\      
${}^*$ The EUF-CMA security proof for {\sf DSA} has not been provided.
Vaudenay \cite{Vau03} mentioned that security of the Brickell model which is a variant of {\sf DSA} can be proven in the ROM.
The Brickell model was presented as an invited talk at CRYPTO 1996 but this model is unpublished.
The construction of the Brickell model and its security proof was  described in \cite{PV96}.
\end{figure*}

\subsection{Road Map}
In Section \ref{Prelimina}, first, we review a digital signature scheme, its security notions.
Next, we review the ROM and WROMs proposed by Numayama et al. \cite{NIT08}.
In Section \ref{OurWROM}, we propose WROMs which capture the chosen prefix collision attack and its variants and give an intuition for simulation method in WROMs.
This simulation technique is needed to prove security in our WROMs.
In Section \ref{Siganalysis}, we analyze the EUF-CMA security against chosen prefix collision attacks for several signature schemes.

In this full version, we provide missing materials in \cite{TYT18}.
In Appendix \ref{SimourWrom}, we provide a simulation method for our WROMs.
In Appendix \ref{security}, we provide missing security proofs for signature schemes.

\section{Preliminaries}\label{Prelimina}
Let $k$ be a security parameter.  A function $f(k)$ is negligible in $k$ if $f(k) \leq 2^{-\omega(\log k)}$.
PPT stands for probabilistic polynomial time.
For strings $m$ and $r$, $|m|$ is the bit length of $m$ and $m||r$ is the concatenation of $m$ and $r$. 
For a finite set $S$, $s \xleftarrow{\$} S$ denotes choosing an element from $S$ uniformly at random. 
For a distribution $\mathcal{D}$, $x \xleftarrow{r} \mathcal{D}$ denotes that $x$ is sampled according to distribution $\mathcal{D}$ and $f_{\mathcal{D}}(x)$ is the probability function of distribution $\mathcal{D}$. 
Let ${\sf B}({N,p})$ be the binomial distribution with $N$ trials and success probability $p$.
The statistical distance of two distributions $P$ and $Q$ over $S$ is defined as $\Delta(P,Q)=\frac{1}{2}\sum_{s \in S}|P(s)-Q(s)|$. 
Let $y \leftarrow \mathcal{A}(x)$ be the output of algorithm $\mathcal{A}$ on input $x$. 

\subsection{Digital Signature Scheme}
We review a digital signature scheme and the EUF-CMA security.
\begin{definition}[Digital Signature Scheme] A digital signature scheme $\Pi$ over the message space $M$ is a triple $\Pi =(\Gen, \Sign, \Verify)$ of PPT algorithms:
\begin{itemize}
 \item[-] \Gen$:$ Given a security parameter $1^k$, return a key pair $(sk, vk)$.
 \item[-] \Sign$:$ Given a signing key $sk$ and a message $m \in M$, return a signature $\sigma$. 
 \item[-] \Verify$:$ Given a verification key $vk$, a message $m$ and a signature $\sigma$, return either 1 $(Accept)$ or 0 $(Reject)$.
\end{itemize}
{\sf Correctness.} For all $k \in \mathbb{N}$, $(sk, vk) \leftarrow \Gen(1^k)$, $m \in M$, we require
\begin{equation*}
\Verify(vk, m, \Sign(sk, m))=1.
\end{equation*}
\end{definition}

\begin{definition}[EUF-CMA]The EUF-CMA security of a digital signature scheme $\Pi$ is defined by the following EUF-CMA game between a challenger $\mathcal{C}$ and PPT adversary $\mathcal{A}$.
\begin{itemize}
 \item[-] $\mathcal{C}$ produces a keypair $(sk,vk) \leftarrow \Gen(1^k)$, and gives the $vk$ to $\mathcal{A}$.
 \item[-] $\mathcal{A}$ makes a number of signing queries $m$ to $\mathcal{C}$. Then, $\mathcal{C}$ signs it as $\sigma \leftarrow  \Sign(sk, m)$, and sends $\sigma$ to $\mathcal{A}$.
 \item[-]  $\mathcal{A}$ outputs a message $m^*$ and its signature $\sigma^*$.  
\end{itemize}
A digital signature scheme satisfies the EUF-CMA security if for all PPT adversaries $\mathcal{A}$, the following advantage of $\mathcal{A}$:
\begin{equation*}
Adv^{\sf EUF\mathchar`-CMA}_{\Pi \mathcal{A}} := \Pr[\Verify(vk, m^*, \sigma^*) = 1  \land m^* {\rm \ is \ not \ queried \ to \ signing}]
\end{equation*}
is negligible in $k$.
\end{definition}

\subsection{Security Notions of Hash Functions}
Let $h: X \rightarrow Y$ be a hash function.
Security notions of a hash function are as follows.
\begin{itemize}
 \item[-]  Collision resistance$:$\\
 It is hard to find a pair $(x, x')$ of inputs such that $h(x) = h(x') \land x \neq x'$ .
  \item[-] Second preimage resistance$:$\\
  Given an input $x \xleftarrow{\$}X$, it is hard to find a second preimage $x' \neq x$ such that $h(x) = h(x')$.
  \item[-] First preimage resistance$:$\\
  Given a hash value $y \leftarrow h(x)$ where $x \xleftarrow{\$}X$, it is hard to find a preimage $x'$ such that $h(x') = y$.
\end{itemize}

\subsection{WROMs Proposed by Numayama et al. \cite{NIT08}}
Let $\ell$ be polynomial in $k$, $X=\{0,1\}^{\ell}$, $Y=\{0, 1\}^k$, $h : X \rightarrow Y$ a random function, and $\mathbb{T}_h = \{ (x, h(x)) \mid x \in X\}$ the table which defines the correspondence between the inputs and outputs of $h$.  To make a more rigorous discussion, we introduce subscript $(\ell, k)$ in the definition,  which make explicit that the length of the input of $h$ is $\ell$ and the length of the output of $h$ is $k$.
Now, we define the random oracle model ROM$_{(\ell, k)}$.

\begin{definition}[ROM$_{(\ell, k)}$]\label{ROM}The random oracle model ROM$_{(\ell, k)}$ is the model that all parties can query the random oracle ${\cal RO}^h$.
\begin{itemize}
 \item[-] {\bf Random oracle ${\cal RO}^h(x)$}\\
 Given an input $x$, the random oracle returns $y$ such that $(x, y)\in\mathbb{T}_h$.
\end{itemize}
\end{definition}
Now, we review the CT-ROM$_{(\ell, k)}$, SPT-ROM$_{(\ell, k)}$ and FPT-ROM$_{(\ell, k)}$ which are defined by Numayama et al. \cite{NIT08}.

\begin{definition}[CT-ROM$_{(\ell, k)}$ \cite{NIT08}]\label{CT-ROM}The collision tractable random oracle model CT-ROM$_{(\ell, k)}$ is the model that all parties can query ${\cal RO}^h$ and the collision oracle ${\cal CO}^h$.

\begin{itemize}
 \item[-] {\bf Collision oracle ${\cal CO}^h()$}\\
 The collision oracle picks one entry $(x, y)\in\mathbb{T}_h$ uniformly at random. If there is any other entry $(x', y)\in\mathbb{T}_h$ then it picks such an entry $(x', y)$ uniformly at random and
 returns $(x, x')$. 
 Otherwise, it returns $\perp$. 
\end{itemize}
\end{definition}
In the CT-ROM$_{(\ell, k)}$, we can capture the situation where the collision resistance property is broken.

\begin{definition}[SPT-ROM$_{(\ell, k)}$ \cite{NIT08}]\label{SPT-ROM}The second preimage tractable random oracle model SPT-ROM$_{(\ell, k)}$ is the model that all parties can query ${\cal RO}^h$ and the second preimage oracle ${\cal SPO}^h$.
\begin{itemize}
 \item[-] {\bf Second preimage oracle ${\cal SPO}^h(x)$}\\
  Given an input $x$, let y be the hash value of $x$ $($i.e., $(x, y) \in \mathbb{T}_h$$)$. If there is any other entry $(x', y)\in\mathbb{T}_h$ such that $x'\neq x$, then  second preimage oracle returns $x'$ uniformly at random.
  Otherwise, it returns $\perp$. 
\end{itemize}
\end{definition}
In the SPT-ROM$_{(\ell, k)}$, we can capture the situation where the second preimage resistance property is broken.

\begin{definition}[FPT-ROM$_{(\ell, k)}$ \cite{NIT08}]\label{FPT-ROM}The first preimage tractable random oracle model FPT-ROM$_{(\ell, k)}$ is the model that all parties can query ${\cal RO}^h$ and the first preimage oracle ${\cal FPO}^h$.
\begin{itemize}
 \item[-] {\bf First preimage oracle ${\cal FPO}^h(y)$}\\
 Given an input $y$, if there is any entry $(x, y)\in\mathbb{T}_h$ then the first preimage oracle returns such $x$ uniformly at random. 
 Otherwise, it returns $\perp$. 
 \end{itemize}
\end{definition}
In the FPT-ROM$_{(\ell, k)}$, we can capture the situation where the preimage resistance property is broken.

\section{WROMs against Chosen Prefix Collision Attacks}\label{OurWROM}
In this section, we propose WROMs which capture the chosen prefix collision attack and its variants.
Let $\ell$ and $t$ be polynomials in $k$, $M=\{0,1\}^{\ell}$, $R=\{0,1\}^{t}$, $X=M\times R$, $Y=\{0, 1\}^k$, $h : X \rightarrow Y$ a random function, and $\mathbb{T}_h = \{ (x, h(x)) \mid x \in X\}$ the table which defines the correspondence between the inputs and outputs of the function $h$.
To make a more rigorous discussion, we introduce subscript $(\ell, t, k)$ in the definition,  which make explicit that the length of the input of the function is $\ell$, the length of the prefix is $t$, and the length of the output is $k$.

\begin{definition}[{\CPCROM}$_{(\ell, t, k)}$]\label{CP-CT-ROM}The common chosen prefix collision tractable random oracle model {\CPCROM}$_{(\ell, t, k)}$ is the model that all parties can query ${\cal RO}^h$ and the common chosen prefix collision oracle ${\cal COMMON\mathchar`-CP\mathchar`-CO}^h$.
\begin{itemize}
 \item[-] {\bf Common chosen prefix collision oracle ${\cal COMMON\mathchar`-CP\mathchar`-CO}^h(r)$}\\
Given an input $r$ $(|r|=t)$, the common chosen prefix collision oracle picks one entry $(m||r, y)\in\mathbb{T}_h$ uniformly at random. If there is any other entry $(m'||r, y)\in\mathbb{T}_h$  then it picks such an entry $(m'||r, y)$ uniformly at random and returns $(m||r, m'||r)$. 
 Otherwise, it returns~$\perp$. 
\end{itemize}
\end{definition}
In the {\CPCROM}$_{(\ell, t, k)}$, we can capture the case of $P=P'$ of the chosen prefix collision attack.

\begin{definition}[{\SPCROM}$_{(\ell, t, k)}$]\label{SP-CT-ROM}The chosen prefix collision tractable random oracle model {\SPCROM}$_{(\ell, t, k)}$ is the model that all parties can query ${\cal RO}^h$ and the chosen prefix collision oracle ${\cal CP\mathchar`-CO}^h$.
\begin{itemize}
 \item[-] {\bf Chosen prefix collision oracle ${\cal CP\mathchar`-CO}^h(r, r')$}\\
 Given an input $(r, r')$ $(|r|=|r'|=t)$, the chosen prefix collision oracle first picks one entry $(m||r, y)\in\mathbb{T}_h$ uniformly at random.
 If there is any other entry $(m'||r', y)\in\mathbb{T}_h$ then it picks such an entry $(m'||r', y)$ uniformly at random and returns $(m||r, m'||r')$. 
  Otherwise, it returns~$\perp$.
\end{itemize}
\end{definition}
In the {\SPCROM}$_{(\ell, t, k)}$, we can capture the chosen prefix collision attack.

\begin{definition}[{\SPSPROM}$_{(\ell, t, k)}$]\label{SP-SPT-ROM}The chosen prefix second preimage tractable random oracle model {\SPSPROM}$_{(\ell, t, k)}$ is the model that all parties can query ${\cal RO}^h$ and the chosen prefix second preimage oracle ${\cal CP\mathchar`-SPO}^h$.
\begin{itemize}
 \item[-] {\bf Chosen prefix second preimage oracle ${\cal CP\mathchar`-SPO}^h(x,r')$}\\
 Given an input $(x, r')$ $(|x|=\ell+t, |r'| = t)$, let y be the hash value of $x$ $($i.e., $(x, y) \in \mathbb{T}_h$$)$. If there is any other entry $(m'||r', y)\in\mathbb{T}_h$ such that $m'||r'\neq x$, then the chosen prefix second preimage oracle returns $m'||r'$ uniformly at random. Otherwise, it returns $\perp$. 
 \end{itemize}
\end{definition}
In the  {\SPSPROM}$_{(\ell, t, k)}$, we can capture a variation of the chosen prefix collision attack that we decide $(P||S, P')$ beforehand and find a collision $(P||S, P'||S')$.

In order to treat the name of GFPT-ROM \cite{TW12} in a similar manner to above models name, we rename it to the {\SPFPROM}$_{(\ell, t, k)}$.

\begin{definition}[{\SPFPROM}$_{(\ell, t, k)}$]\label{SP-FPT-ROM}The chosen prefix first preimage tractable random oracle model {\SPFPROM}$_{(\ell, t, k)}$ is the model that all parties can query ${\cal RO}^h$ and the chosen prefix first preimage oracle ${\cal CP\mathchar`-FPO}^h$.

\begin{itemize}
 \item[-] {\bf Chosen prefix first preimage oracle ${\cal CP\mathchar`-FPO}^h(y,r)$}\\
 Given an input $(y,r)$ $(|y|=k, |r| = t)$, if there is an entry $(m||r, y)\in\mathbb{T}_h$ then the chosen prefix first preimage oracle returns such $m||r$ uniformly at random. 
 Otherwise, it returns $\perp$. 
\end{itemize}
\end{definition}

By Definition \ref{CP-CT-ROM}, \ref{SP-CT-ROM}, \ref{SP-SPT-ROM}, \ref{SP-FPT-ROM}, following relations among WROMs hold. 
\begin{itemize}
 \item[-] If security of a cryptographic scheme is proven in the {\SPCROM}$_{(\ell, t, k)}$(resp., {\SPSPROM}$_{(\ell, t, k)}$, {\SPFPROM}$_{(\ell, t, k)}$), then the scheme satisfies this security in the {\CPCROM}$_{(\ell, t, k)}$ (resp., {\SPCROM}$_{(\ell, t, k)}$, {\SPSPROM}$_{(\ell, t, k)}$).
\end{itemize}

\subsubsection*{Intuition of Simulation for WROMs.}
In the ROM, the reduction algorithm simulates ${\cal RO}^h$ using a table $\mathbb{T}$ which has entries $(x, y)$ representing that the hash value of $x$ is $y$. 
In WROMs, the reduction algorithm must simulate the additional oracle.
In the CT-ROM$_{(\ell, k)}$, SPT-ROM$_{(\ell, k)}$ and FPT-ROM$_{(\ell, k)}$,  the behavior of the additional oracle depends on the number of preimages. 
The reduction algorithm uses $\mathbb{T}$ and a table $\mathbb{L}$.
The table $\mathbb{L}$ has entries $(y, n)$ representing that $y$ has $n$ preimages. 
In the {\CPCROM}$_{(\ell, t, k)}$, {\SPCROM}$_{(\ell, t, k)}$, {\SPSPROM}$_{(\ell, t, k)}$ and {\SPFPROM}$_{(\ell, t, k)}$, the behavior of the additional oracle depends on not only the number of preimages but also prefixes.
The reduction algorithm simulates the additional oracle by adding an entry for prefix $r$ to tables $\mathbb{T}$ and $\mathbb{L}$.
Concretely, $\mathbb{T}$ has entries $((m,r), y)$ representing that a hash value of $m||r$ is $y$ and $\mathbb{L}$ has entries $((y,r), n)$ representing that $y$ has $n$ preimages which have the prefix $r$.
We will describe technical details in Appendix \ref{SimourWrom}.

\section{Security of Signature Schemes in WROMs}\label{Siganalysis}

In this section, we argue the EUF-CMA security of signature schemes in WROMs. 
We will describe the proof of security analyses of signature schemes in Appendix \ref{security}.
Before analyzing RSA based signature schemes, we recall the RSA assumption.

\begin{definition}[RSA Generator]The RSA generator \RSAGen, which on input $1^k$, randomly chooses distinct $k/2$-bit primes $p$, $q$ and computes $N=pq$ and $\phi=(p-1)(q-1)$. It randomly picks $e \xleftarrow{\$} \mathbb{Z}_{\phi(N)}$ and computes $d$ such that $ed =1 \ {\rm mod}\ \phi(N)$. The RSA generator outputs $(N, e, d)$.
\end{definition}

\begin{assumption}[RSA Assumption]\label{RSAassum}A polynomial-time machine $\mathcal{A}$ is said to solve the RSA problem if given an RSA instance $(N, e, z)$ where $N$, $e$ are generated by {\sf RSA}$(1^k)$ and $z \xleftarrow{\$} \mathbb{Z}^{*}_{N}$, it outputs $z^{1/e}\ {\rm mod}\ N$ with non-negligible probability.
The RSA assumption is that there is no PPT adversary that solves the RSA problem.
\end{assumption}

\subsection{RSA-FDH}

Let $\ell$ be a polynomial in $k$, $M=\{0, 1\}^{\ell}$ the message space, and $h: \{0, 1\}^{\ell} \rightarrow \{0, 1\}^{k}$ a hash function. {\sf RSA-FDH} \cite{BR96} is described in Fig. \ref{FDH}.

\begin{figure*}[htbp]
  \begin{center}
    \begin{tabular}{|l|l|l|} \hline
      \Gen$(1^k)$                         & \Sign$(sk, m)$              & \Verify$(vk, m, \sigma)$ \\ \hline
      $(N, e, d) \leftarrow \RSAGen(1^k) $   & $y \leftarrow h(m)$              & $y \leftarrow \sigma^e$ mod $N$ \\
      $vk \leftarrow (N, e)$                                         & $\sigma \leftarrow y^d$ mod $N$  & if $h(m)=y$ \\
      $sk \leftarrow (N, d)$                & return $\sigma$                  & \ \ \ return 1 \\
      return $(vk, sk)$                &                                  & else \\                            &                                  & \ \ \ return 0 \\  \hline                                
    \end{tabular}   
    \label{tab:const}
  \end{center}
   \caption{{\sf RSA-FDH}}\label{FDH}
\end{figure*}

\begin{theorem}\label{FDH-CP-CT}In the {\CPCROM}$_{(\ell_1, t_1,k)}$ $(\ell_1+t_1=\ell)$, there exists a PPT adversary $\mathcal{A}$ that breaks {\sf RSA-FDH} by making queries to the signing oracle and ${\cal COMMON\mathchar`-CP\mathchar`-CO}^h$ with probability at least $1-e^{(1-2^{\ell_1})/2^k}$.
\end{theorem}
Security proof of Theorem \ref{FDH-CP-CT} is given in Appendix \ref{FDH-INSECURE}.

\subsection{RSA-PFDH}
Let $\ell$ and $k_1$ be polynomials in $k$, $M=\{0, 1\}^{\ell}$ the message space, and $h: \{0, 1\}^{\ell + k_1} \rightarrow \{0, 1\}^{k}$ a hash function. {\sf RSA-PFDH} \cite{Cor02} is described  in Fig. \ref{PFDH}.

\begin{figure*}[htbp]
  \begin{center}
    \begin{tabular}{|l|l|l|} \hline
      \Gen$(1^k)$                                 & \Sign$(sk, m)$                  & \Verify$(vk, m, \sigma)$ \\ \hline
      $(N, e, d) \leftarrow \RSAGen(1^k)$  & $r \xleftarrow{\$} \{0, 1\}^{k_1}$     & parse $\sigma$ as $(r, x)$ \\
      $vk \leftarrow (N, e)$                         & $y \leftarrow h(m||r)$                & $y \leftarrow x^e$ mod $N$ \\
      $sk \leftarrow (N, d)$                         & $x \leftarrow y^d$ mod $N$       & if $h(m||r)=y$\\
      return $(vk, sk)$                                & $\sigma \leftarrow (r, x)$        & \ \ \ return 1  \\ 
                                                                &  return $\sigma$                                    & else  \\
                                                                &                & \ \ \ return 0 \\  \hline                                               
    \end{tabular}
    \label{tab:const}
  \end{center}
        \caption{{\sf RSA-PFDH}}\label{PFDH}

\end{figure*}

\begin{theorem}\label{PFDH-SP-CT}In the {\SPCROM}$_{(\ell_1, t_1, k)}$ $(\ell_1 + t_1 = \ell + k_1)$, for all PPT adversaries $\mathcal{B}$ that break {\sf RSA-PFDH} with probability $\epsilon_{\rm euf}$ by making $q_{sign}$, $q_h$ and $q_{sc}$ queries to the signing oracle, ${\cal RO}^h$, and ${\cal CP\mathchar`-CO}^h$ respectively. 
There exists a PPT adversary $\mathcal{A}$ that solves the RSA problem with $\epsilon_{\rm rsa}$ such that 
\begin{equation*}
\epsilon_{\rm euf} \leq \epsilon_{\rm rsa} + \frac{1}{2^k} + \frac{q_{sign}Q_2}{2^{k_1}} + Q_1 \times p_{{\bf prefixRO}^h} 
\end{equation*}
where $Q_1 = q_{sign} + q_h +q_{sc} + 1$, $Q_2 = q_{sign} + q_h +2q_{sc} + 1$ and
\begin{equation*}
p_{{\bf prefixRO}^h}\leq 
\begin{cases}
    \frac{\ln2^k}{\ln\ln2^k}\frac{10Q_2}{2^k} +\frac{1}{(2^k)^2}+\frac{Q_2}{2^k}& (\ell_1 \geq k) \\
    \frac{\ln2^k}{\ln\ln2^k}\frac{10Q_2}{2^{\ell_1}} +\frac{1}{(2^k)^2}+\frac{Q_2}{2^k} & (\ell_1 < k). \\    
      \end{cases}
\end{equation*} 
\end{theorem}
Security proof of Theorem \ref{PFDH-SP-CT} is given in Appendix \ref{PFDH-SECURE}.

\begin{theorem}\label{PFDH-SP-SPT}In the {\SPSPROM}$_{(\ell, k_1, k)}$, there exists a PPT adversary $\mathcal{A}$ that breaks {\sf RSA-PFDH} by making queries to the signing oracle and ${\cal CP\mathchar`-SPO}^h$ with probability at least $1-e^{(1-2^{\ell})/2^k}$. If $\ell \geq k \geq 2$, $\mathcal{A}$ outputs a valid forgery with probability at least $1-e^{-1/2}$.
\end{theorem}
Security proof of Theorem \ref{PFDH-SP-SPT} is given in Appendix \ref{PFDH-INSECURE}.

\subsection{RSA-PFDH$^\oplus$}

Let $\ell$ be a polynomial in $k$, $M=\{0, 1\}^{\ell}$ the message space, and $h: \{0, 1\}^{\ell + k} \rightarrow \{0, 1\}^{k}$ a hash function. 
{\sf RSA-PFDH}$^\oplus$ \cite{NIT08} is described  in Fig. \ref{PFDHo+}.

\begin{figure*}[htbp]
  \begin{center}
    \begin{tabular}{|l|l|l|} \hline
      \Gen$(1^k)$                                 & \Sign$(sk, m)$         & \Verify$(vk, m, \sigma)$ \\ \hline
      $(N, e, d) \leftarrow \RSAGen(1^k)$  & $r \xleftarrow{\$} \{0, 1\}^{k}$ & parse $\sigma$ as $(r, x)$ \\
      $vk \leftarrow (N, e)$                         & $w \leftarrow h(m||r)$      & $y \leftarrow x^e$ mod $N$ \\
      $sk \leftarrow (N, d)$                         & $y \leftarrow w\oplus r$    & $w \leftarrow h(m||r)$ \\
      return $(vk, sk)$                                 & $x \leftarrow y^d$ mod $N$      & if $w\oplus r = y$ \\ 
                                                                & $\sigma \leftarrow (r, x)$    & \ \ \ return 1 \\
                                                                & return $\sigma$               & else \\      
                                                                 &                               & \ \ \ return 0 \\  \hline
    \end{tabular}
    \label{tab:const}
  \end{center}
     \caption{{\sf RSA-PFDH}$^\oplus$}\label{PFDHo+}
\end{figure*}

\begin{theorem}\label{PFDHo+-SP-CT}In the \SPCROM$_{(\ell_1, t_1, k)}$ $(\ell_1 + t_1 = \ell + k)$, for all PPT adversaries $\mathcal{B}$ that break {\sf RSA-PFDH}$^\oplus$ with probability $\epsilon_{\rm euf}$ by making $q_{sign}$, $q_h$ and $q_{sc}$ queries to the signing oracle, ${\cal RO}^h$, and ${\cal CP\mathchar`-CO}^h$, respectively. There exists a PPT adversary $\mathcal{A}$ that solves the RSA problem with $\epsilon_{\rm rsa}$ such that 
\begin{equation*}
\epsilon_{\rm euf} \leq \epsilon_{\rm rsa} + \frac{1}{2^k} + \frac{q_{sign}Q_2}{2^{k_1}} + Q_1 \times p_{{\bf prefixRO}^h}
\end{equation*}
where $Q_1 = q_{sign} + q_h +q_{sc}+ 1$, $Q_2 = q_{sign} + q_h +2q_{sc}+ 1$ and
\begin{equation*}
p_{{\bf prefixRO}^h}\leq 
\begin{cases}
    \frac{\ln2^{k}}{\ln\ln2^{k}}\frac{10Q_2}{2^{k}} +\frac{1}{(2^{k})^2}+\frac{Q_2}{2^{k}}& (\ell_1 \geq k) \\
    \frac{\ln2^{k}}{\ln\ln2^{k}}\frac{10Q_2}{2^{\ell_1}} +\frac{1}{(2^{k})^2}+\frac{Q_2}{2^{k}} & (\ell_1 < k). \\    
      \end{cases}
\end{equation*} 
\end{theorem}
Security proof of Theorem \ref{PFDHo+-SP-CT} is given in Appendix \ref{PFDHo+-SECURE}.

\begin{theorem}\label{PFDHo+-SP-SPT}In the {\SPSPROM}$_{(\ell, k, k)}$, there exists a PPT adversary $\mathcal{A}$ that breaks {\sf RSA-PFDH}$^\oplus$ by making queries to the signing oracle and the chosen prefix second preimage oracle for $h$ with probability at least $1-e^{(1-2^\ell)/2^k}$.
\end{theorem}
Security proof of Theorem \ref{PFDHo+-SP-SPT} is given in Appendix \ref{PFDHo+-INSECURE}.

\subsection{RSASSA-PKCS-v1.5}
We discuss {\sf RSASSA-PKCS-v1.5} \cite{PKCS}. To simplify the discussion, exclude detailed settings and treat octet strings as binary strings.
Let $\ell$, $k_1$, and $k_2$ be polynomials in $k$, and $c$ a constant which is determined by the type of hash function when implementing {\sf RSASSA-PKCS-v1.5}. Let $M=\{0, 1\}^\ell$ be the message space and $h: \{0, 1\}^{\ell} \rightarrow \{0, 1\}^{k_2}$ a hash function.
Let $s$ be a fixed binary string of length $k_1$. The $HashAlgID$ represent the type of hash function when implementing {\sf RSASSA-PKCS-v1.5} in a specific binary string of length $c$.
The $HashAlgID$ and the string $s$ are published in advance.
Let $k=k_1+c+k_2$.
{\sf RSASSA-PKCS-v1.5} is described  in Fig. \ref{RSASSA}.

\begin{figure*}[htbp]
  \begin{center}
    \begin{tabular}{|l|l|l|} \hline
      \Gen$(1^k)$                               & \Sign$(sk, m)$              & \Verify$(vk, m^*, \sigma^*)$ \\ \hline
      $(N, e, d) \leftarrow \RSAGen(1^k)$          & $w \leftarrow h(m)$              & $y^* \leftarrow (\sigma^*)^e$ mod $N$ \\
      $vk \leftarrow (N, e)$                         & $y \leftarrow s||HashAlgID||w$   & parse $y^*$ as $s^*||HashAlgID^*||w^*$ \\
      $sk \leftarrow (N, d)$                         & $x \leftarrow y^d$ mod $N$       & if $s^*||HashAlgID^* \neq s||HashAlgID$\\
      return $(vk, sk)$                              & $\sigma \leftarrow x$            & \ \ \ return 0 \\ 
                                                     & return $\sigma$                  & if $h(m^*) = w^*$\\
                                                     &                                  & \ \ \ return 1 \\      
                                                     &                                  & else \\
                                                     &                                  & \ \ \ return 0 \\  \hline
                                         
    \end{tabular}
    \label{tab:const}
  \end{center}
     \caption{{\sf RSASSA-PKCS-v1.5}}\label{RSASSA}
\end{figure*}

\begin{theorem}\label{RSASSAPKCS-CT}In the CT-ROM$_{(\ell, k_2)}$, there exists a PPT adversary $\mathcal{A}$ that breaks {\sf RSASSA-PKCS-v1.5} by making queries to the signing oracle and ${\cal CO}^h$ with probability at least $1-e^{(1-2^\ell)/2^{k_2}}$.
\end{theorem}

\begin{theorem}\label{RSASSAPKCS-CP-CT}In the {\CPCROM}$_{(\ell_1, t_1, k_2)}$ $(\ell_1 + t_1 = \ell)$, there exists a PPT adversary $\mathcal{A}$ that breaks {\sf RSASSA-PKCS-v1.5} by making queries to the signing oracle and ${\cal COMMON\mathchar`-CP\mathchar`-CO}^h$ with probability at least $1-e^{(1-2^{\ell_1})/2^{k_2}}$.
\end{theorem}

Theorem \ref{RSASSAPKCS-CT} and \ref{RSASSAPKCS-CP-CT} can be proven similar way as in Theorem \ref{FDH-CP-CT}.

\subsection{DSA}
We discuss {\sf DSA} \cite{Ker13}. We recall a group generator algorithm \GrGen.
\begin{definition}[Group Generator]The group generator algorithm \GrGen, which on input $1^k$, randomly chooses a $k$-bit prime $q$  and a $j$-bit prime such that  $q | (p-1)$ $(j$ is polynomial in $k)$. 
It chooses an $x \in \{1, ... ,p-1\}$ such that $x^{(p-1)/q} \not\equiv 1$ mod $p$ and sets $g = x^{(p-1)/q}$ mod $p$.
The DSA setup algorithm outputs $(p, q, g)$.
\end{definition}

Let $\ell$ be a polynomial in $k$, $M=\{0, 1\}^{\ell}$ the message space, and $h: \{0, 1\}^{\ell + k} \rightarrow \{0, 1\}^{k}$ a hash function.
{\sf DSA} is described  in Fig. \ref{DSA}.

\begin{figure*}[htbp]
  \begin{center}
    \begin{tabular}{|l|l|l|} \hline
      \Gen$(1^k)$                               & \Sign$(sk, m)$                           & \Verify$(vk, m, \sigma)$ \\ \hline
   $(p, q, g) \leftarrow \GrGen(1^k)$    & $k \xleftarrow{\$} [0,q-1]$                   & parse $\sigma$ as $(r, s)$\\
   $x \xleftarrow{\$} \{0, ... ,q-1\} $             & $r \leftarrow (g^k {\rm mod}\ p)$ mod $q$     & $w \leftarrow s^{-1}$ mod $q$\\     
   $y \leftarrow g^x$                               & $z \leftarrow h(m)$                           & $z \leftarrow h(m)$ \\ 
   $vk \leftarrow (p, q, g, y)$                  & $s \leftarrow (k^{-1}(z+xr))$ mod $q$         & $u_1 \leftarrow zw$  mod $q$,  $u_2 \leftarrow rw$  mod $q$\\
   $sk \leftarrow (p, q, g, x)$                  & $\sigma \leftarrow (r, s)$                    & $v \leftarrow g^{u_1}y^{u_2}$  mod $q$\\                              
   return $(vk, sk)$                                 & return $\sigma$                               & if $v = r$\\ 
                                                     &                                               &\ \ \ return 1   \\ 
                                                     & &else\\ 
                                                     & &\ \ \ return 0\\ \hline                                                                                                                                         
    \end{tabular}
    \label{tab:const}
     \caption{{\sf DSA}}\label{DSA}
  \end{center}
\end{figure*}

 \begin{theorem}\label{DSA-CT}In the CT-ROM$_{(\ell+k, k)}$, there exists a PPT adversary $\mathcal{A}$ that breaks {\sf DSA} by making queries to the signing oracle and ${\cal CO}^h$ with probability at least $1-e^{(1-2^{\ell+ k})/2^k}$.
\end{theorem}

\begin{theorem}\label{DSA-CP-CT}In the {\CPCROM}$_{(\ell_1, k_1, k)}$ $(\ell_1 + k_1 = \ell+k)$, there exists a PPT adversary $\mathcal{A}$ that breaks {\sf DSA} by making queries to the signing oracle and ${\cal COMMON\mathchar`-CP\mathchar`-CO}^h$ with probability at least $1-e^{(1-2^{\ell_1})/2^k}$.
\end{theorem}

Theorem \ref{DSA-CT} and \ref{DSA-CP-CT} can be proven similar way as in Theorem \ref{FDH-CP-CT}.

\section{Conclusion}
In this paper, we analyze the security of standard signature schemes against chosen prefix collision attacks by defining three WROMs. 
Our analysis of these schemes in WROMs provides a more precise security indication against chosen prefix attacks.
We showed {\sf RSA-PFDH} and {\sf RSA-PFDH}$^\oplus$ are EUF-CMA secure in the {\SPCROM}, but not secure in the {\SPSPROM}. 
We also showed that {\sf RSASSA-PKCS-v1.5} and {\sf DSA} are not EUF-CMA secure in both the CT-ROM and {\CPCROM}.

When discussing the security in the {\SPCROM}$_{(\ell, t, k)}$, we fixed the length of the prefix $\ell$. 
Studying the case of variable length is a future work. 
There are practical signature schemes and encryption schemes which have not been analyzed in WROMs.
The security analysis of these schemes in WROMs is also an interesting future work.

\smallskip
\noindent
\section*{Acknowledgement.}
A part of this work was supported by Input Output Hong Kong, Nomura Research Institute, NTT Secure Platform Laboratories, Mitsubishi Electric, I-System, JST CREST JPMJCR14D6, JST OPERA, and JSPS KAKENHI 16H01705, 17H01695.
We are grateful to Kazuo Ohta (University of Electro-Communications) and Shiho Moriai (National Institute of Information and Communications Technology) for giving us the opportunity to do this research.
We would also like to thank anonymous referees for their constructive comments.

\bibliographystyle{abbrvurl}
\bibliography{refW}

\appendix 

\section{Simulation Method in the CP-CT-ROM}\label{SimourWrom}

We propose a simulation method in the {\SPCROM}$_{(\ell, t, k)}$ based on a simulation method of the {\sf CP-FPT-ROM}$_{(\ell, t, k)}$ which is proposed by Tan and Won \cite{TW12}.
We describe the algorithm {\preRO} and {\CPCO} which simulate ${\cal RO}^h$ and ${\cal CP\mathchar`-CO}^h$ in the {\SPCROM}$_{(l, t, k)}$.
Since a behavior of ${\cal CP\mathchar`-CO}^h$ depends on the number of preimages and prefixes, we use two tables.
{\preRO} and {\CPCO} share two tables $\mathbb{T}$, $\mathbb{L}$, which are empty in their initial state.
The table $\mathbb{T}$ has entries $((m,r), y)$ representing that a hash value of $m||r$ is $y$.
The table $\mathbb{L}$ has entries $((y,r), n)$ representing the number of $m$ such that $m||r$ is a preimage of $y$ for some $m$.
The two algorithms perform the simulation while synchronizing the table $\mathbb{T}$ and $\mathbb{L}$ so that there is no inconsistency between them. Let $\#\mathbb{T}$ be the number of entries of $\mathbb{T}$, $\#\mathbb{L}$ the number of entries of $\mathbb{L}$, $\#\mathbb{T}(r')$ the number of entries of $((m, r'),y) \in \mathbb{T}$ for some $m$ and $y$, $\#\mathbb{T}(r',y')$ the number of entries of $((m, r'),y') \in \mathbb{T}$ for some $m$, and $\#\mathbb{L}(r')$ the number of entries of $((y, r'),n)$ for some $y$ and $n$.

\begin{lemma}\label{pre.ballbin}Let $X = M \times R $, $h: X \rightarrow Y$ a hash function, $\#Y \geq 2$, and $n_{y, r}$ represent the number of preimages of $y$ that satisfy the conditions of $h(x)=y$ and $x = m||r$ for some $m \in M$ under a function $h$.
Let ${\sf BAD}_r$ be the event that there is some $y$ such that $n_{y,r} > L$ where if $\#M \geq \#Y$, $L = \frac{5 \ln \#Y}{\ln\ln \#Y} \frac{\#M}{\#Y}$, or otherwise $L = \frac{5 \ln \#Y}{\ln\ln \#Y}$.

 \begin{equation*}
\Pr[{\sf BAD}_r] < \frac{1}{(\#Y)^2}
  \end{equation*}
\end{lemma}
Lemma \ref{pre.ballbin} is obtained by letting $X$ of Lemma 1 in \cite{KNTX10} correspond to $M \times \{r\}$.

\begin{lemma}\label{simpre.RO}The distribution on the outputs of ${\cal RO}^h$ is equal to the distribution on the outputs of the algorithm {\preRO}.
\end{lemma}
Lemma \ref{simpre.RO} is an extension of Lemma 1 in \cite{NIT08}.

\begin{lemma}\label{simpre.WROM}The distribution on the outputs of ${\cal RO}^h$ and ${\cal CP\mathchar`-CO}^h$ is equal to the distribution on the outputs of algorithms {\preRO} and {\CPCO}.
\end{lemma}
Lemma \ref{simpre.WROM} is an extension of Corollary 1 in \cite{NIT08}. 

\begin{lemma}\label{unipre.WROM}
Let $X=M\times R$ and $h: X \rightarrow Y$ a hash function in {\SPCROM}$_{(\ell, t, k)}$.
Let $\mathcal {A}$ be a PPT oracle query machine that queries $q_h$ and $q_{sc}$ times for ${\cal RO}^h$ and ${\cal CP\mathchar`-CO}^h$ respectively and $q = q_h + 2q_{sc}$. 
Let $H_{\mathcal{A},h}(x)$ be a random variable that represents a hash value ${\cal RO}^h(x)$, where $x = m||r \leftarrow \mathcal{A}^{{\cal RO}^h,{\cal CP\mathchar`-CO}^h}$ and the correspondence $(x, h(x))$ is not returned by the two oracles. Then any $\mathcal {A}$, the following inequality holds where $2q \leq \#M$ and $2q \leq \#Y$:
\begin{equation*}
\begin{split}
\Delta(H_{\mathcal{A},h}(x), U_Y)\leq 
\begin{cases}
    \frac{1}{\#Y}\left(5q + 1 + \frac{4q^2}{\#Y} + 20q\frac{\ln \#Y}{\ln\ln\#Y}\right) & (\#M \geq \#Y) \\
    \frac{1}{\#M}\left(5q + 1 + \frac{4q^2}{\#M} + 20q\frac{\ln \#Y}{\ln\ln\#Y}\right) & (\#M < \#Y) .\\    
      \end{cases}
\end{split}
\end{equation*}
\end{lemma}
Lemma \ref{unipre.WROM} is an extension of Lemma 2 in \cite{KNTX10}. 

\begin{lemma}\label{propre.WROM}Let $h$ be a hash function in the {\SPCROM}$_{(\ell, t, k)}$ and {\bCPCO} the part of  {\preRO} of the algorithm changed to {\bpreRO}. Let $q_h$ and $q_{sc}$ be the number of queries to ${\cal RO}^h$ and ${\cal CP\mathchar`-CO}^h$ respectively and $q= q_h + 2q_{sc}$.
Except for the following probability $p_{{\bf prefixRO}^h}$, the distributions of two tables $\mathbb{L}$ and $\mathbb{T}$ used in the simulation ${\bf RO}^h$ and \CPCO are identical to two tables $\mathbb{L}$ and $\mathbb{T}$ used in  the simulation {\bpreRO} and {\bCPCO}.
\begin{equation*}
p_{{\bf prefixRO}^h}\leq 
\begin{cases}
    \frac{\ln\#Y}{\ln\ln\#Y}\frac{10q}{\#Y} +\frac{1}{(\#Y)^2}+\frac{q}{\#Y}& (\#M \geq \#Y) \\
    \frac{\ln\#Y}{\ln\ln\#Y}\frac{10q}{\#M} +\frac{1}{(\#Y)^2}+\frac{q}{\#Y} & (\#M < \#Y) \\    
      \end{cases}
\end{equation*} 
\end{lemma}

\vspace{0pt}
\begin{algorithm}[htbp]
  \caption{{\preRO}$(x)$}\label{pre.RO}
  \begin{algorithmic}[1]
    \State Parse $x$ as $m||r$.
    \State If there is an entry $((m, r), y) \in \mathbb{T}$ for some $y$, then return $y$.
    \State Compute the following value:
    \begin{equation*}
		p = \frac{\sum_{((\tilde{y}, r), \tilde{n}) \in \mathbb{L}}(\tilde{n} - \#\mathbb{T}(r,\tilde{y}))} {\#M - \#\mathbb{T}(r)}.
    \end{equation*} 
    \State Flip a biased coin with $\Pr[\alpha = 0 ] = p$.
    \State If $\alpha = 0$, then return $y$ as follows.     
    \begin{itemize}
        \item[(a)] Pick $y \xleftarrow{r} \mathcal{D}$ according to the following distribution:
    \begin{equation*}
       f_{\mathcal{D}}(y) = \frac{n - \# \mathbb{T}(r, y)} {\sum_{((\tilde{y}, r), \tilde{n}) \in \mathbb{L}}(\tilde{n} - \#\mathbb{T}(r, \tilde{y}))}\ \  for\ ((y, r), n)\in\mathbb{L}.
    \end{equation*}
    \item[(b)] Insert $((m, r), y)$ in $\mathbb{T}$ and return $y$.
    \end{itemize}    
    \State If $\alpha = 0$, then return $y$ as follows. 
    
    \begin{itemize}
 \item[(a)] Pick $y \xleftarrow{\$} Y \setminus   \left( \bigcup_{((\tilde{y}, r), \tilde{n}) \in \mathbb{L}} \tilde{y}\right)$ uniformly at random. 
 \vspace{8pt}
 \item[(b)] 
$n' \xleftarrow{r} {\sf B}\Bigl(\#M-\sideset{}{_{((\tilde{y}, r), \tilde{n}) \in \mathbb{L}}}\sum\tilde{n}  -1, \frac{1}{\#Y - \# \mathbb{L}(r)}\Bigl)$.
 \vspace{8pt}
\item[(c)]  $n \leftarrow n'+1$. 
\item[(d)] Insert $((y, r), n)$ in $\mathbb{L}$, insert $((m, r), y)$ in $\mathbb{T}$, and return $y$.
 \end{itemize}  
      \end{algorithmic}
\end{algorithm}    
\vspace{0pt}

\begin{algorithm}[htbp]
  \caption{{\bf CP-CO}$^h(r, r')$}\label{SP-CO}
  \begin{algorithmic}[1]
    \State Pick uniformly $m \xleftarrow{\$} M$.
    \State $x \leftarrow m||r$.
    \State Run the algorithm {\preRO}$(x)$ and get the hash value $y = h(m||r)$.
    \State If there is no entry $((y, r'), n) \notin \mathbb{L}$ for any $n$, then
    \begin{itemize}
       \item[(a)] 
  $n' \xleftarrow{r} {\sf B}\Bigl(\#M-\sideset{}{_{((\tilde{y}, r), \tilde{n}) \in \mathbb{L}}}\sum\tilde{n} , \frac{1}{\#Y - \# \mathbb{L}(r)}\Bigl)$.
      \item[(b)]   Insert $((y, r'), n')$ in $\mathbb{L}$.    
    \end{itemize}
   \State   If $((y, r'), 0) \in \mathbb{L}$, then return $\perp$.
   \State If $r = r'$
    \begin{itemize}
    \item[(a)] If $((y, r), 1) \in \mathbb{L}$, then return  $\perp$.
    \item[(b)] Compute the following value:
    \begin{equation*}
q_{((y, r), n)}=\frac{\#\mathbb{T}(r, y)-1}{n-1}.
  \end{equation*}
    \item[(c)] Flip a biased coin with $\Pr[\beta = 0 ] = q_{((y, r), n)}$.
    \item[(d)] If $\beta =0$ 
    \begin{itemize}
    \item[(i)] Pick uniformly one entry $((m',r), y) \in \mathbb{T}$ satisfying $m \neq m'$. 
    \item[(ii)] Return $(m||r, m'||r)$.
\end{itemize}
    \item[(e)] If $\beta =1$
\begin{itemize}
 \item[(i)] Pick uniformly $m' \leftarrow M$ such that there is no entry $((m', r), \tilde{y}) \in \mathbb{T}$ for any $\tilde{y} \in Y$.
 \item[(ii)] Insert $((m', r), y)$ in $\mathbb{T}$ and return $(m||r, m'||r)$.
  \end{itemize}    
    \end{itemize}
   \State If $r \neq r'$
   \begin{itemize}
    \item[(a)] Compute the following value:
    \begin{equation*}
q_{((y, r), n)}=\frac{\#\mathbb{T}(r', y)}{n}.
  \end{equation*}
    \item[(b)] Flip a biased coin with $\Pr[\beta = 0 ] = q_{((y, r), n)}$.
    \item[(c)] If $\beta =0$
    \begin{itemize}
 \item[(i)] Pick uniformly one entry $((m',r'), y) \in \mathbb{T}$ satisfying $m \neq m'$. 
 \item[(ii)] Return $(m||r, m'||r')$.
\end{itemize}
    \item[(d)] If $\beta =1$
\begin{itemize}
 \item[(i)] Pick uniformly $m' \leftarrow M$ such that there is no entry $((m', r'), \tilde{y}) \in \mathbb{T}$ for any $\tilde{y} \in Y$. 
 \item[(ii)] Insert $((m', r'), y)$ in $\mathbb{T}$ and return $(m||r, m'||r')$.
 \end{itemize}    
    \end{itemize}
      \end{algorithmic}
\end{algorithm}
\vspace{0pt}

\begin{algorithm}[htbp]
  \caption{{\bpreRO}$(x)$}
  \begin{algorithmic}[1]
    \State Parse $x$ as $m||r$.
    \State If there is an entry $((m, r), y) \in \mathbb{T}$ for some $y$, then return $y$.
    \State $y \xleftarrow{\$} Y$.
    \State If $y \in \displaystyle {\bigcup_{((\tilde{y}, r), \tilde{n}) \in \mathbb{L}}} \tilde{y}$ then abort.
    \State $n' \xleftarrow{r} {\sf B}\Bigl(\#M-\sum_{((\tilde{y}, r), \tilde{n}) \in \mathbb{L}}\tilde{n}  -1, \frac{1}{\#Y - \# \mathbb{L}(r)}\Bigl)$.
    \State $n \leftarrow n'+1$. 
    \State Insert $((y, r), n)$ in $\mathbb{L}$, insert $((m, r), y)$ in $\mathbb{T}$, and return $y$.
      \end{algorithmic}
\end{algorithm}    
\vspace{0pt}

\begin{proof}
Let {\sf E} be the event that step 5 in {\preRO} does not occur and {\sf F} be the event that step 3 in {\bpreRO} is $y \notin \bigcup_{((\tilde{y}, r), \tilde{n}) \in \mathbb{L}} \tilde{y}$.
If both {\sf E} and {\sf F} occur, The behavior of {\preRO} and {\bpreRO} is identical. 
\begin{equation*}
\begin{split}
\Pr[\lnot  {\sf E}]&=\frac{\sum_{((\tilde{y}, r), \tilde{n}) \in \mathbb{L}}(\tilde{n} - \#\mathbb{T}(r,\tilde{y}))} {\#M - \#\mathbb{T}(r)} \leq \frac{\sum_{((\tilde{y}, r), \tilde{n}) \in \mathbb{L}}\tilde{n} } {\#M - \#\mathbb{T}(r)}
\end{split}
\end{equation*}
Consider the case where {\sf BAD}$_r$ does not occur. 
\begin{equation*}
\begin{split}
\Pr[\lnot  {\sf E}| \lnot  {\sf BAD}_r]&=\frac{qL}{\#M - \#\mathbb{T}(r)} =\frac{qL}{\#M}\left(\frac{1}{1-\frac{\#\mathbb{T}(r)}{\#M}}\right) \leq \frac{qL}{\#M}\left(1+ \frac{\#\mathbb{T}(r)}{\#M}\right) \leq \frac{2qL}{\#M}.
\end{split}
\end{equation*}
where if $\#M \geq \#Y$, $L = \frac{5 \ln \#Y}{\ln\ln \#Y} \frac{\#M}{\#Y}$, otherwise $L = \frac{5 \ln \#Y}{\ln\ln \#Y}$\\
The probability $\Pr[\lnot  {\sf F}]$ evaluates follows.
\begin{equation*}
\begin{split}
\Pr[\lnot  {\sf F}] \leq \frac{q}{\#Y} 
\end{split}
\end{equation*}
By Lemma \ref{pre.ballbin}, we have
\begin{equation*}
\begin{split}
\Pr[\lnot ( {\sf E} \land  {\sf F})]& \leq \Pr[\lnot  {\sf E}] + \Pr[\lnot  {\sf F}] \\
& \leq \Pr[\lnot  {\sf E} | \lnot  {\sf BAD}_r] + \Pr[ {\sf BAD}_r] + \Pr[\lnot  {\sf F}] \\
& \leq \frac{2qL}{\#M} + \frac{1}{(\# Y)^2} + \frac{q}{\#Y}.
\end{split}
\end{equation*}
Hence
\begin{equation*}
\begin{split}
p_{{\bf prefixRO}^h} &= \Pr[\lnot ( {\sf E} \land  {\sf F})]\\
& \leq \Pr[\lnot  {\sf E} | \lnot  {\sf BAD}_r] + \Pr[ {\sf BAD}_r] + \Pr[\lnot  {\sf F}] \\
& \leq \begin{cases}
    \frac{\ln\#Y}{\ln\ln\#Y}\frac{10q}{\#Y} +\frac{1}{(\#Y)^2}+\frac{q}{\#Y}& (\#M \geq \#Y) \\
    \frac{\ln\#Y}{\ln\ln\#Y}\frac{10q}{\#M} +\frac{1}{(\#Y)^2}+\frac{q}{\#Y} & (\#M < \#Y). \\    
      \end{cases}
\end{split}
\end{equation*}\qed
\end{proof}

\section{Security Proof for Signature Schemes}\label{security}
\subsection{Proof of Theorem \ref{FDH-CP-CT}}\label{FDH-INSECURE}

\begin{proof}We construct an algorithm $\mathcal{A}$ as follows.
\begin{itemize}
 \item[(1)] Query ${\cal COMMON\mathchar`-CP\mathchar`-CO}^h$ with $p$ and obtain $\xi$.
 \item[(2)] If $\xi =\perp$ then abort, otherwise parse $\xi$ as $(m||p, m'||p)$,  where $h(m||p) = h(m'||p)$.
 \item[(3)] Query the signature of $m||p$ to the signing oracle, and obtain a signature $\sigma$.
 \item[(4)] Output $(m'||p, \sigma)$ as a valid forgery.
 \end{itemize}
 If $\mathcal{A}$ does not abort, then $\mathcal{A}$ can output a valid forgery. Let {\sf abort} be the event that $\mathcal{A}$ aborts.
 
\begin{equation*}
\begin{split}
\Pr \left[ {\sf abort} \right]
& = \Pr\left[ \xi = \perp  \right ] \\
& = \Pr\left[ \# \mathbb{T}_h(y, r) \leq 1 \right ] \\
& =\Pr\left[ \# \mathbb{T}_h(y, r) = 1 \right ] \\
& = \Pr\left[ n' = 0 | n'  \xleftarrow{r} {\sf B} \left( 2^{\ell_1}-1, \frac{1}{2^k} \right) \right ] \\
& = \left( 1 - \frac{1}{2^k}  \right)^{2^{\ell_1}-1} \leq e^{(1-2^{\ell_1})/2^k}\\
\end{split}
\end{equation*}
Therefore, $\mathcal{A}$ can output a valid forgery with probability at least $1-e^{(1-2^{\ell_1})/2^k}$. \qed
\end{proof}

\subsection{Proof of Theorem \ref{PFDH-SP-CT}}\label{PFDH-SECURE}

\begin{proof}Assume that  a PPT algorithm $\mathcal{B}$ breaks the EUF-CMA security with $\epsilon_{\rm euf}$ which is non-negligible in $k$. To prove the theorem, we first describe a sequence of games.  Let {\bf  Game} $0$ be the original {\sf  EUF-CMA} game in the {\SPCROM}$_{(\ell_1, t_1, k)}$ and {\bf Game} $6$ be directly related to solve the RSA problem. Let $S_i$ be the event that an adversary outputs a valid forgery in the {\bf Game} $i$.

\begin{itemize}
 \item {\bf  Game} $0$: The original EUF-CMA game in the {\SPCROM}$_{(\ell_1, t_1, k)}$.
  \begin{equation*}
 \Pr[S_0]=\epsilon_{\rm euf}
 \end{equation*}
 \item {\bf  Game} $1$: We replace ${\cal RO}^h$ and ${\cal CP\mathchar`-CO}^h$  by algorithms {\preRO} and {\CPCO} respectively. 
 Let tables $\mathbb {T}$ and $\mathbb {L}$ be simulation tables commonly used in algorithms {\preRO} and {\CPCO}.
 By Lemma \ref{simpre.WROM}, we have
  \begin{equation*}
 |\Pr[S_0]-\Pr[S_1]|=0.
 \end{equation*}
 \item {\bf  Game} $2$: We replace algorithms {\preRO} and {\CPCO} by algorithms {\bpreRO} and {\bCPCO}. {\bCPCO} is the part of  {\preRO} in {\CPCO} changed to {\bpreRO}. $ Q_1 $ is the total number of queries to {\preRO}. By Lemma \ref{propre.WROM}, we have
 \begin{equation*}
 |\Pr[S_1]-\Pr[S_2]| \leq Q_1 \times p_{{\bf prefixRO}^h}.
 \end{equation*}
 \item {\bf Game} $3$: When the signing algorithm runs {\bpreRO} on input $m||r$, parse $m||r$ as $b||c$ $(|b|=\ell_1, |c| = t_1)$. If there is an entry $((b, c),y) \in \mathbb{T}$ for some $y$ already, then {\bpreRO} aborts. $Q_2$ is the bound of the number of entries recorded in the table $\mathbb{T}$. 
  \begin{equation*}
  |\Pr[S_2]-\Pr[S_3]| \leq \frac{q_{sign}Q_2}{2^{k_1}}
  \end{equation*}
 \item {\bf Game} $4$: For the setting of a hash value of {\bpreRO}, fix $z \leftarrow \mathbb{Z}_N^*$ and change it as follows.
 \begin{itemize}
 \item If the hash value is queried by the signing algorithm, then {\bpreRO} chooses $x \in \mathbb{Z}_N$ and outputs $y = x^e$ mod $N$.
 \item If the hash value is queried by the adversary or {\bCPCO}, then {\bpreRO} chooses $x \in \mathbb{Z}_N$ and outputs $y = zx^e$ mod $N$. 
\end{itemize}

\begin{equation*}
|\Pr[S_3]-\Pr[S_4]|=0
\end{equation*}
 \item {\bf  Game} $5$: We modify the signing algorithm in the computation $y^d$ to search $(x, y)$ such that $x = y^d$ instead of using the signing key $d$. 
 \begin{equation*}
 |\Pr[S_4]-\Pr[S_5]|=0
 \end{equation*}
  \item {\bf Game} $6$: When receiving the output forgery $(m^*, \sigma^*)$ from the adversary, parse $\sigma^*$ as $\sigma^* = r^*|| x^*$.  If $m^*||r^*$  is not queried to {\bpreRO}, then aborts. 
 \begin{equation*}
 |\Pr[S_5]-\Pr[S_6]|\leq \frac{1}{2^k}
 \end{equation*}
\end{itemize}

We construct the algorithm $\mathcal{A}$ which breaking the RSA assumption using the algorithm $\mathcal{B}$. The operation of $\mathcal{A}$ for the input RSA instance $(N, e, z^*)$ is changed to the $z$ in {\bf Game} 6 to $z^*$. Suppose $\mathcal{A}$ do not abort receiving a forgery $ (m^*, \sigma^*)$ from $\mathcal{B}$. When parsing $\sigma^*$ as $r^*||x^*$, then $h(m^*||r^*)=y^* =(x^*)^e$ holds. When $\mathcal{A}$ computes $ (z^*) ^ {1/e} = x^*/x$ using $x$ chosen by {\bpreRO} for the query of the hash value of $m||r$ from $\mathcal {B}$ then $(x^*/x) ^ e = z^*$ holds.
Hence, $\mathcal{A}$ can output the solution $(z^*)^{1/e}$ of the RSA instance $(N, e, z^*)$. We can bound the probability $\Pr[S_6] \leq \epsilon_{\rm rsa}$.

\begin{equation*}
\epsilon_{\rm euf} \leq \epsilon_{\rm rsa} + \frac{1}{2^k} + \frac{q_{sign}Q_2}{2^{k_1}} + Q_1 \times p_{{\bf prefixRO}^h}
\end{equation*}
Therefore, $\mathcal{A}$ breaks the RSA assumption with non-negligible probability $\epsilon_{\rm rsa}$. \qed
\end{proof}

\subsection{Proof of Theorem \ref{PFDH-SP-SPT}}\label{PFDH-INSECURE}
\begin{proof}We construct an algorithm $\mathcal{A}$ as follows.
\begin{itemize}
 \item[(1)] Query the signature of $m$ to the signing oracle, and obtain a signature $\sigma$.
 \item[(2)] Parse $\sigma$ as $(r, x)$.
 \item[(3)] Query ${\cal CP\mathchar`-SPO}^h$ with $((m||r), r)$, and obtain $\xi$.
 \item[(4)] If $\xi =\perp$ then abort, otherwise parse $\xi$ as $m'||r$ where $h(m||r) = h(m'||r)$.
 \item[(5)] Output $(m', \sigma)$ as a valid forgery.
 \end{itemize}
 If $\mathcal{A}$ does not abort, then $\mathcal{A}$ can output a valid forgery. Let {\sf abort} be the event that $\mathcal{A}$ aborts.
 
\begin{equation*}
\Pr \left[ {\sf abort} \right]
 = \Pr\left[ \# \mathbb{T}_h(y, r) = 1 \right ] 
 \leq e^{(1-2^{\ell})/2^k}
\end{equation*}
Therefore, $\mathcal{A}$ can output a valid forgery with probability at least $1-e^{(1-2^{\ell})/2^k}$. \qed
\end{proof}

\subsection{Proof of Theorem \ref{PFDHo+-SP-CT}}\label{PFDHo+-SECURE}
\begin{proof} We modify the {\bf Game} $4$ in the proof of Theorem \ref{PFDH-SP-CT} as follows.

 \item {\bf Game} $4$: For the setting of a hash value of {\bpreRO}, fix $z \leftarrow \mathbb{Z}_N^*$ and change it as follows.
 \begin{itemize}
 \item If the hash value is queried by the signing algorithm, then {\bpreRO} chooses $x \in \mathbb{Z}_N$ and  $y = x^e$ mod $N$, then outputs $w=y \oplus r$.
 \item If the hash value is queried by the adversary or {\bCPCO}, then {\bpreRO} chooses $x \in \mathbb{Z}_N$ and $y = zx^e$ mod $N$, then outputs $w=y \oplus r$.
\end{itemize}
It can be shown in a similar way as in Theorem \ref{PFDH-SP-CT}. \qed
\end{proof}

\subsection{Proof of Theorem \ref{PFDHo+-SP-SPT}}\label{PFDHo+-INSECURE}
\begin{proof}We construct an algorithm $\mathcal{A}$ as follows.
\begin{itemize}
 \item[(1)] Query the signature of $m$ to the signing oracle, and obtain a signature $\sigma$.
 \item[(2)] Parse $\sigma$ as $(r, x)$.
 \item[(3)] Query ${\cal CP\mathchar`-SPO}^h$ with $((m||r), r)$, and obtain $\xi$.
 \item[(4)] If $\xi =\perp$ then abort, otherwise parse $\xi$ as $m'||r$ where $h(m||r) = h(m'||r)$.
 \item[(5)] Output $(m', \sigma)$ as a valid forgery.
 \end{itemize}
 If $\mathcal{A}$ does not abort, then $\mathcal{A}$ can output a valid forgery. Let {\sf abort} be the event that $\mathcal{A}$ aborts.
 
\begin{equation*}
\Pr \left[ {\sf abort} \right]
 = \Pr\left[ \# \mathbb{T}_h(y, r) = 1 \right ] 
 \leq e^{(1-2^{\ell})/2^k}
\end{equation*}
Therefore, $\mathcal{A}$ can output a valid forgery with probability at least $1-e^{(1-2^{\ell})/2^k}$. \qed
\end{proof}

\setcounter{tocdepth}{2}
\tableofcontents

\end{document}